\documentclass[english,aps,floats,twocolumn,showpacs,prl,amsfonts,nofootinbib]{revtex4}

\usepackage{pslatex,amssymb,amsmath,amsfonts,graphicx,wasysym}
\usepackage[T1]{fontenc}
\usepackage[latin1]{inputenc}

\newcommand{\mpl}{M_{\rm Pl}}

\begin{document}

\vspace*{-10ex}
\hspace*{\fill}
MAD-TH-08-16

\title{Towards a warped inflationary brane scanning}

\author{Heng-Yu Chen}
 \email{hchen46@wisc.edu}
 \affiliation{Department of Physics, University of Wisconsin-Madison, Madison, WI
 53706-1390, USA}

\author{Jinn-Ouk Gong}
 \email{jgong@lorentz.leidenuniv.nl}
 \affiliation{Instituut-Lorentz for Theoretical Physics, Universteit Leiden, 2333 CA
 Leiden, The Netherlands}

\begin{abstract}

We present a detailed systematics for comparing warped brane inflation with the
observations, incorporating the effects of both moduli stabilization and ultraviolet
bulk physics. We explicitly construct an example of the inflaton potential governing
the motion of a mobile D3 brane in the entire warped deformed conifold. This allows
us to precisely identify the corresponding scales of the cosmic microwave
background. The effects due to bulk fluxes or localized sources are parametrized
using gauge/string duality. We next perform some sample scannings to explore the
parameter space of the complete potential, and first demonstrate that without the
bulk effects there can be large degenerate sets of parameters with observationally
consistent predictions. When the bulk perturbations are included, however, the
observational predictions are generally spoiled. For them to remain consistent, the
magnitudes of the additional bulk effects need to be highly suppressed.

\end{abstract}


\maketitle

\section{Constructing a potential for warped brane inflation}

The inflationary paradigm~\cite{inflation} addresses a number of fine tuning problems of
the standard hot big bang cosmology, such as the horizon and the flatness problems. It
also predicts a nearly scale invariant power spectrum of the curvature perturbation,
which has been verified to high accuracies by the observation of the thermal fluctuations
in the cosmic microwave background (CMB) and the large scale structure of the
universe~\cite{Tegmark:2006az,Komatsu:2008hk}. Numerous models of inflation based on
effective field theory have been proposed, however, distinct predictions of a given model
crucially depend on its ultraviolet completion. To construct a truly predictive
inflationary model, it is clearly important to embed it into a consistent microscopic
theory of quantum gravity such as string theory.

During the past few years, our understanding of the various ingredients for obtaining
string inflation has been significantly expanded, and many models with increasing
sophistication and striking signatures have been proposed (For recent developments, see
Ref.~\cite{Reviews} and references therein). In the coming decade, {beyond the ongoing
Sloan Digital Sky Survey~\cite{Tegmark:2006az} and the Wilkinson Microwave Anisotropy
Probe~\cite{Komatsu:2008hk}, vastly improved cosmological data will become available from
the advanced CMB observations~\cite{Planck}, the CMB polarization
experiments~\cite{cmbpolarization}, the dark energy surveys~\cite{jdem} as well as the
map of large scale structure~\cite{euclid}.} They will allow us to constrain the
parameter spaces of these models, and possibly to even rule out some of them. It is
therefore of timely interest to perform a thoroughly updated and complete case study in
such a direction\footnote{What we mean by ``complete'' should become clear momentarily.}.
In this paper we shall focus on one of the most developed string inflation models in the
literature, usually referred to as ``brane inflation''.

The original setup of brane inflation first introduced in Ref.~\cite{DvaliTye} is to
consider a pair of spacetime-filling D3-$\overline{\rm D3}$ branes, separated at some
distance greater than the local string length on a compact six manifold. As D3 and
$\overline{\rm D3}$ move towards each other under the Coulombic attraction, the
canonical inflaton is then identified as the separation between them. Unfortunately
in such a simple setup, the Coulombic attraction is too strong for the slow-roll
inflation to persist. To overcome this obstacle, the authors of Ref.~\cite{KKLMMT}
considered instead placing the $\rm D3$-$\overline{\rm D3}$ pair in a locally warped
deformed conifold throat developed in a compact Calabi-Yau orientifold by background
fluxes. The $\overline{\rm D3}$ is then stabilized at the tip of deformed conifold, and
D3 is attracted weakly by the warped down D3-$\overline{\rm D3}$ potential given by
\begin{equation}\label{VD3D3bar}
V_{D3\overline{D3}}(\tau,\sigma) = \frac{D_0}{U^2} \left( 1 - \frac{3D_0}{16\pi^2
T_3^2{|\bf{y}-\bar{\bf{y}}|}^4} \right) \, .
\end{equation}
Here $D_0=2T_3a_0^4$, $T_3=1/[(2\pi)^3 g_s(\alpha')^2]$ is the D3 brane tension,
$a_0=\exp[{-2\pi K/(3g_sM)}]$ is the warp factor at the tip of deformed conifold
with $-K$ and $M$ being the quanta of NSNS and RR three form fluxes, and
$|\bf{y}-\bar{\bf{y}}|$ is the D3-$\overline{\rm D3}$ separation. Furthermore, the
factor $U$ is the universal K\"ahler modulus, whose role we shall discuss
in detail later. Notice that the first term in (\ref{VD3D3bar}) gives the positive
contribution and uplifts the total potential energy, whereas the Coulombic
attraction is highly suppressed by $a_0^8$ and only becomes dominating near the tip
of deformed conifold. Eventually when $|\bf{y}-\bar{\bf{y}}|$ becomes comparable to
warped string length $\sim a_0 l_s$, D3 and $\overline{\rm D3}$ annihilate through
open string tachyon condensation, rapidly terminates the brane inflation.

There are two important further ingredients that have so far been missing in our
discussion, and they are crucial for obtaining the inflationary phase and making detailed
comparisons with observational data. The first ingredient is the stabilization of both
closed and open sting moduli. They are usually stabilized by the perturbative flux
potential~\cite{GVW} and the non-perturbative superpotential generated by wrapped
branes~\cite{npW}. The second ingredient is the ultraviolet corrections arising from
embedding the warped throat into a compact Calabi-Yau orientifold. The bulk fluxes and
the distant branes, or additional supersymmetry breaking and moduli stabilization sources
can give significant perturbations to the inflaton potential derived from the local
sources and geometry.

The warped deformed conifold offers us an ideal venue to analyze these two
ingredients. Its explicit metric~\cite{Conifold} allows for studying the moduli
stabilization and the construction of the inflaton potential valid for the {\em
entire} evolution, including precise identification of where inflation ends.
Furthermore while bulk physics is largely unknown, the spectrum of supergravity
states in singular conifold has been tabulated in Ref.~\cite{T11}. The gauge/string
duality then allows us to parametrize these symmetry breaking bulk perturbations to
the inflaton potential by coupling dual (approximately) conformal field theory
to these bulk modes~\cite{Baumann2}. Combining these with the D3-$\overline{\rm D3}$
interaction, we can schematically parametrize the total potential of the inflaton
$\phi$ including both local and bulk effects, experienced by a mobile D3 brane in
the warped deformed conifold as
\begin{equation}\label{TotalV}
{\mathbb{V}}(\phi) = V_{D3\overline{D3}}(\phi) + V_{\rm stab.}(\phi) + V_{\rm
bulk}(\phi) \, .
\end{equation}
Here $V_{\rm stab.}(\phi)$ arises from moduli stabilization and $V_{\rm bulk}(\phi)$
encodes all other possible perturbations from bulk physics\footnote{Although we follow
similar scheme as in Ref.~\cite{Baumann2}, we do not partition $V_{\rm stab.}(\phi)$ such
that the inflaton mass term $\sim H^2\phi^2$ is singled out. As $H^2$ is usually a
combination of microscopic parameters, for the purpose of full parameter scanning we
shall calculate $V_{\rm stab.}(\phi)$ in full detail and express it explicitly in terms
of the microscopic parameters while treating $V_{\rm bulk}(\phi)$ as further
perturbations.}. Most of the quantities specifying $V_{\rm stab.}(\phi)$ are exclusively
related to the local geometry of the throat, e. g. the warp factor $a_0$. However $V_{\rm
stab.}(\phi)$ typically also depends on other quantities controlled by bulk physics, such
as the one loop determinant of the non-perturbative superpotential. The quantities
controlling $V_{\rm stab.}(\phi)$ are usually treated as free parameters and yield a
landscape of possible inflaton potentials.

The current observational data~\cite{Tegmark:2006az,Komatsu:2008hk} enable us to make
comparisons with the predictions yielded by different parameter sets and constrain their
allowed values. In order for this exercise to be instructive, it is crucial to include
all the significant contributions to the inflaton potential. The existing literature in
this direction~\cite{branescan,ClineScan} has mostly focused on the first two
contributions in (\ref{TotalV}) without taking into account $V_{\rm bulk}(\phi)$. However
in light of recent results in Ref.~\cite{Baumann2} indicating that $V_{\rm bulk}(\phi)$
can be generically comparable to $V_{\rm stab.}(\phi)$, it is clearly necessary to apply
such general results and scan the enlarged parameter space to include the bulk
corrections.

Here we aim to provide some initial steps toward a complete systematic parameter scanning
for warped brane inflation. We shall first consider a specific brane configuration and
stabilize explicitly both universal K\"ahler modulus and some of the angular moduli of
D3. We then construct an example of $V_{D3\overline{D3}}(\phi)+V_{\rm stab.}(\phi)$ valid
for {\em entire} warped deformed conifold throat. This potential should be regarded as
the infrared completion to the model obtained in Ref.~\cite{Baumann1} under the singular
conifold limit (See also \cite{McGill1,KP} for related work). It allows us to identify
the end point of inflation, hence extrapolate precisely to the CMB scale. Next we shall
briefly review the parametrization of the bulk effects $V_{\rm bulk}(\phi)$ given in
Ref.~\cite{Baumann2}, and discuss the microscopic and observational constraints on the
inflationary parameter scanning. Finally we shall first present different degenerate
parameter sets such that the resultant $V_{D3\overline{D3}}(\phi)+V_{\rm stab.}(\phi)$
yields observationally consistent curvature power spectrum $\mathcal{P}_\mathcal{R}$ and
the corresponding spectral index $n_\mathcal{R}$. We next demonstrate that the
perturbations due to $V_{\rm bulk}(\phi)$ can have significant impact and need to be
small to preserve these seemingly optimal parameter sets. These sample scannings aim to
highlight the possible degeneracies and the important role of bulk effects.

\section{Enumerating the inflaton potential}

\subsection{Moduli stabilization potential from warped throat}

In this section, we shall explicitly consider the effects of moduli stabilization on
the mobile D3 brane in the entire deformed conifold. This is important for
accurate comparisons between the predictions and the observational data. In
particular, this allows us to identify {\em precisely} the end point of inflation
$\phi_{\rm e}$, defined to be the point where the slow-roll parameter
\begin{equation}\label{SRepsilon}
\varepsilon \equiv 2\mpl^2 \left[ \frac{H'(\phi)}{H(\phi)} \right]^2 \approx
\frac{\mpl^2}{2} \left[ \frac{{\mathbb{V}}'(\phi)}{{\mathbb V}(\phi)} \right]^2
\end{equation}
becomes 1 so that the universe ceases accelerated expansion. Here, $\mpl=(8\pi
G)^{-1/2}$, $H(\phi)$ is the field dependent Hubble parameter and a prime denotes a
derivative with respect to $\phi$. Note that the second approximation holds under the
slow-roll limit. It is crucial to properly take into account the late evolution of the
universe during inflation for making correct inflationary predictions. The form of the
potential near the end of inflation can substantially lower the inflationary energy
scale~\cite{German:2001tz}, and the light fields other than the canonical inflaton can
completely dominate the curvature power spectrum $\mathcal{P}_\mathcal{R}$ and the
corresponding spectral index $n_\mathcal{R}$~\cite{curvaton}. Furthermore, the
post-inflationary evolution can also modify the spectral index at an observationally
detectable level~\cite{Choi:2008et}.

In the context of warped brane inflation, $\varepsilon(\phi)$ is driven towards $1$
mostly by the D3-$\overline{\rm D3}$ Coulombic attraction, which is exponentially
suppressed and only becomes significant inside the deformed conifold. Moreover, as shown
in Ref.~\cite{CGS1} when Coulombic attraction is insignificant, $\varepsilon(\phi)$
remains small all the way to the tip for generic parameter sets. We therefore expect
inflation persists well into the deformed conifold region, despite only a proportionally
small number of $e$-folds is expected to be generated there. Moreover as some of the
inflationary predictions are already tightly constrained by observations to high degrees
of accuracies, e.g. $\mathcal{P}_\mathcal{R}$ and $n_\mathcal{R}$, it is important to
take into account such infrared completion in constraining the parameter space of brane
inflation.

The key component capturing the moduli stabilization effects on the mobile D3 is
the ${\mathcal{N}}=1$ supergravity $F$-term scalar potential,
\begin{equation}\label{DefVF}
V_{F}(z^\alpha,\bar{z}^\alpha,\rho,\bar{\rho}) = e^{\kappa^2{\cal K}} \left( {\cal
K}^{\Sigma\Omega}D_{\Sigma}W\overline{D_{\Omega}W} - 3\kappa^2|W|^2 \right) \, ,
\end{equation}
where $\kappa^2=\mpl^{-2}$. Let us discuss various contributions to (\ref{DefVF}) in
turn. In the presence of a D3 brane, the universal K\"ahler modulus $U(z,\rho)$
depends on the brane position $\{z^\alpha\,,\bar{z}^\alpha\}$ and the usual complex
bulk K\"ahler modulus $\rho=\sigma+i\chi$. The indices in (\ref{DefVF}) therefore
runs over $\{\rho, z^{\alpha}\}$ and total K\"ahler potential is the given
by~\cite{DG}
\begin{align}\label{eq:KahlerPotential}
\kappa^2 {\cal K}( z^\alpha, \bar{z}^{\alpha},\rho,\bar{\rho}) = & -3 \log \left[
\rho+\bar{\rho}-\gamma k \left( z^\alpha,\bar{z}^\alpha \right) \right]
\nonumber\\
\equiv & -3 \log U(z,\rho) \, .
\end{align}
Here, $k(z^\alpha,\bar{z}^\alpha)$ is the geometric K\"ahler potential of the metric on
the Calabi-Yau, and $\gamma = \sigma_0T_3/(3M_P^2)$ is the normalization constant with
$\sigma_0$ being the value of $\sigma$ when the D3 brane is at its stabilized
configuration~\cite{Baumann1}, including the uplifting potentials\footnote{Strictly
speaking, the derivation of (\ref{eq:KahlerPotential}) given in Ref.~\cite{DG} is invalid
for the warped background, hence raises the question about the validity of
(\ref{eq:KahlerPotential}) itself in the warped deformed conifold. However, some
interesting new development in Ref.~\cite{WarpKahlerPotential} about the universal
K\"ahler modulus indicates that (\ref{eq:KahlerPotential}) can remain valid in the warped
background. It would be useful to verify this by combining the earlier work on the
dynamics of warped compactification, e.g. Ref.~\cite{GM} with the recent results,
Ref.~\cite{WarpKahlerPotential}. We thank Bret Underwood for discussing with us about
this issue.}.

To stabilize some of the geometric and K\"ahler moduli, we need to consider the total
superpotential $W(z^\alpha,\rho)$ consisting of two contributions as
\begin{equation}\label{eq:SuperW}
W(z^\alpha,\rho) =  W_0 + A(z^{\alpha})e^{-a \rho} \, .
\end{equation}
The first term $W_0=\int G_3\wedge \Omega_3$ is the perturbative flux
superpotential~\cite{GVW}, which allows us (at least in principle) to stabilize the
complex structure moduli and dilaton-axion. One mechanism for stabilizing $\rho$ and
some of the mobile D3 brane position moduli is to include non-perturbative effects
through gaugino condensation on a stack of space-filling D7 branes (or a Euclidean
D3 brane instanton), as appears in the second term of (\ref{eq:SuperW}). The
prefactor $A(z^\alpha)$ is a holomorphic function of the D3 brane moduli and can
be written as~\cite{Wprefactor}
\begin{equation}\label{DefAz}
A(z^\alpha) = A_0 \left[ \frac{f(z^\alpha)}{f(0)} \right]^{1/n} \, .
\end{equation}
Here $A_0$ is a complex constant whose exact value depends on the stabilized complex
structure moduli, and $n$ is the number of D7 (or $n=1$ for Euclidean D3) giving
the gaugino condensate (or instanton correction). The parameter $a$ in
(\ref{eq:SuperW}) is given by $2\pi/n$. The explicit dependence on the position of
mobile D3 brane appears through the holomorphic embedding function $f(z^\alpha) =
0$ of the four cycle in the Calabi-Yau space.

Substituting the total superpotential (\ref{eq:SuperW}) as well as the expression
for the inverse metric ${\cal K}^{\Sigma\Omega}$ solved in Ref.~\cite{McGill1} into
(\ref{DefVF}), the explicit form of $V_F(z^\alpha,\bar{z}^{\alpha},\rho,\bar{\rho})$
is given by
\begin{align}\label{explicitVF}
& V_F(z^\alpha,\bar{z}^\alpha,\rho,\bar{\rho})
\nonumber\\
&= \frac{\kappa^2}{3[U(z,\rho)]^2} \left\{ \left[ U(z,\rho) + \gamma
k^{\gamma\bar\delta} k_\gamma k_{\bar{\delta}} \right] |W_{,\rho}|^2 - 3 \left(
\overline{W}W_{,\rho} + c.c. \right) \right\}
\nonumber\\
& \hspace{0.3cm} + \frac{\kappa^2}{3[U(z,\rho)]^2} \left[ \left(
k^{\alpha\bar\delta}k_{\bar\delta}\overline{W}_{,\bar{\rho}}W_{,\alpha} + c.c.
\right) + \frac{1}{\gamma} k^{\alpha\bar\beta}W_{,\alpha}\overline{W}_{,\bar{\beta}}
\right] \, .
\end{align}
Here, the subscript of a letter with a comma denotes a partial differentiation with
respect to the corresponding component. Specifically for a deformed conifold defined by
the complex embedding equation $\sum_{\alpha=1}^4(z^\alpha)^2=\epsilon^2$ with $z^\alpha
\in {\mathbb{C}}$, the K\"ahler potential is given by
\begin{equation}\label{DefconKahlerpotential}
k(\tau) = \frac{\epsilon^{4/3}}{2^{1/3}} \int_\tau d\tau' \left[ \sinh(2\tau') -
2\tau' \right]^{1/3} \, .
\end{equation}
In writing (\ref{DefconKahlerpotential}), we have also used the standard relation
$\sum_{\alpha=1}^4|z^\alpha|^2=\epsilon^2\cosh\tau$ (See Refs.~\cite{Conifold,KS}
for the explicit metric in terms of $\tau$ and angular coordinates). To apply the
general formula (\ref{explicitVF}), we note the inverse metric $k^{\bar{i}j}$ is
given by
\begin{equation}\label{InverseMetric2}
k^{\bar{i}j} = \frac{r^3}{k''} \left[ R^{\bar{i}j} + \coth\tau \left( \frac{k''}{k'} -
\coth\tau \right) L^{\bar{i}j} \right] \, ,~~(i,\bar{j}=1,2,3)
\end{equation}
where $k' = dk/d\tau$ and $k'' = d^2k/d\tau^2$, and the $3\times 3$ matrices
${R}^{\bar{i}{j}}$ and ${L}^{\bar{i}{j}}$ in (\ref{InverseMetric2}) are,
respectively,\footnote{Here, we have made the substitution
$z_4=\pm\sqrt{\epsilon^2-(z_1^2+z_2^2+z_3^2)}$.}
\begin{align}\label{defRijLij}
R^{\bar{i}j} = & \delta^{\bar{i}j} - \frac{z_i\bar{z}_j}{r^3} \, ,
\\
L^{\bar{i}j} = & \left( 1 - \frac{\epsilon^4}{r^6} \right) \delta^{\bar{i}j} +
\frac{\epsilon^2}{r^3} \frac{z_iz_j + \bar{z}_i\bar{z}_j}{r^3} - \frac{z_i\bar{z}_j
+ \bar{z}_iz_j}{r^3} \, .
\end{align}
We can now readily calculate various terms depending on the inverse deformed conifold
metric $k^{\bar{i}j}$ in the $F$-term scalar potential (\ref{explicitVF}). First we
notice that $L^{\bar{i}j}$ has the property
$k_{\bar{i}}{{L}}^{\bar{i}{j}}={{L}}^{\bar{i}{j}}k_j=0$; therefore, the norm
$k^{\bar{i}j}k_{\bar{i}}k_j$ is given by
\begin{equation}\label{normkk}
k^{\bar{i}j}k_{\bar{i}}k_j = \frac{3}{4} \frac{\epsilon^{4/3}}{2^{1/3}}\frac{ \left[
\sinh(2\tau) - 2\tau \right]^{4/3}}{\sinh^2\tau} \, .
\end{equation}
Similarly, we can calculate that
\begin{widetext}
\begin{align}
\label{normkW}
k^{\bar{i}j}k_{\bar{i}}W_j = & \frac{3}{4} \frac{\cosh\tau}{\sinh^3\tau} \left[
\sinh(2\tau)-2\tau \right] \sum_{j=1}^3 \left( z_j - \bar{z}_j
\frac{\epsilon^2}{r^3} \right) A_j e^{-a\rho} \, ,
\\
k^{\bar{i}j}\overline{W}_{\bar{i}}W_j = & \frac{3}{2\cdot2^{2/3}} \epsilon^{2/3}
\frac{\cosh\tau}{\sinh\tau^2} \left[ \sinh(2\tau) - 2\tau \right]^{2/3} \left\{
R^{\bar{i}j}\overline{W}_{\bar{i}}W_j + \left[ \frac{2}{3}
\frac{\sinh(2\tau)}{\sinh(2\tau)-2\tau} - \coth^2\tau \right] \times
L^{\bar{i}j}\overline{W}_{\bar{i}}W_j \right\} \, ,
\end{align}
where
\begin{align}
\label{normRWW}
R^{\bar{i}{j}}\overline{W}_{\bar{i}}W_j = & e^{-2a\sigma} \left[ \sum_{i=1}^3
|A_i|^2 - \frac{1}{r^3} \left( \sum_{i=1}^3 z_i \bar{A}_i \right) \left(
\sum_{j=1}^3 \bar{z}_j {A}_j \right) \right] \, ,
\\
\label{normLWW}
L^{\bar{i}{j}}\overline{W}_{\bar{i}}W_j = & e^{-2a\sigma} \left\{ \sum_{i=1}^3 \left( 1 -
\frac{\epsilon^4}{r^6} \right)|A_i|^2 - \frac{1}{r^3} \left( \sum_{i,j=1}^3 \bar{A}_i
\left[ z_i\bar{z}_j + z_j\bar{z}_i - \frac{\epsilon^2}{r^3} \left( z_iz_j +
\bar{z}_i\bar{z}_j \right) \right]{A}_j \right) \right\} \, .
\end{align}
\end{widetext}
Putting various components together, we can obtain the general expression of the
$F$-term scalar potential in deformed conifold. We shall see that for a specific
D7 embedding given in Ref.~\cite{Kuperstein}, the resultant expression nicely
simplifies along its angular stable trajectory.

\subsection{A case study}

As an explicit example, we consider specifically the D7 brane embedding given
by~\cite{Kuperstein}
\begin{equation}\label{Kuperembedding}
f(z^{\alpha})=z_1-\mu \, ,
\end{equation}
from which we can easily find that
\begin{align}\label{AKuper}
A(z^\alpha) = & A_0 \left( 1 - \frac{z_1}{\mu} \right)^{1/n} \, ,
\\
A_i(z^\alpha) = & -\frac{A_0}{n\mu} \left( 1 - \frac{z_1}{\mu} \right)^{1/n-1} \delta_{i1}
\, .
\end{align}
Without lost of generality, we shall take $\mu\in {\mathbb{R}}^{+}$. The embedding
(\ref{Kuperembedding}) is highly symmetric, and preserves $SO(3)$ subgroup of the
full $SO(4)$ continuous isometry group of the deformed conifold.\footnote{The actual
angular stable trajectory however only preserves $SO(2)$ subgroup of $SO(3)$.}
Substituting $A_i(z^\alpha)$ and $\overline{{A}_j(z^\alpha)}$ into the earlier
expressions derived for the $F$-term scalar potential, the dependence on the D3
brane position now only appears through the combinations $(z_1+\bar{z}_1)$ and
$|z_1|^2$. The resultant expression therefore has the functional dependence $V_F =
V_F\left(z_1+\bar{z}_1,|z_1|^2,\tau,\sigma,\chi\right)$.

In addition, we also need to stabilize some of the moduli appearing in
$V_F\left(z_1+\bar{z}_1,|z_1|^2,\tau,\sigma,\chi\right)$ following the standard
procedure outlined in Ref.~\cite{Baumann1}. First the axion of the complex K\"ahler
modulus $\chi$ can be stabilized by rotating the phase of the flux induced
superpotential $W_0 \in {\mathbb{R}}^{-}$, and making the replacement
$\exp(ia\chi)/A(z^\alpha) \to 1/|A(z^\alpha)|$. As the isometry of the deformed
conifold is partially broken by D7 branes, some of the angular coordinates of the
mobile D3 can also be stabilized by the resultant $F$-term scalar potential. In
Ref.~\cite{CGS1}, such specific angular stable trajectory for the D7 embedding
(\ref{Kuperembedding}) for the entire deformed conifold is derived to be
\begin{equation}\label{StabTraj}
z_1 = -\epsilon \cosh \left( \frac{\tau}{2} \right) \, .
\end{equation}
We refer the readers to Ref.~\cite{CGS1} for the derivation of this trajectory and
the discussion about its stability\footnote{Furthermore, as the angular dependences
are only encoded in the $F$-term scalar potential (at least for the region where
most of $e$-folds occur), we expect including the uplifting term does not affect the
stability analysis.}. The resultant two-field scalar potential $V_F(\tau,\sigma)$,
for such an angular stable trajectory, is thus given by
\begin{align}\label{Vfts}
V_F(\tau,\sigma) = & \frac{2a^2\kappa^2|A_0|^2e^{-2a\sigma}}{[U(\tau,\sigma)]^2}
|g(\tau)|^{2/n}
\nonumber\\
& \times \left\{ \frac{U(\tau,\sigma)}{6} + \frac{1}{a} \left( 1 -
\frac{|W_0|}{|A_0|} \frac{e^{a\sigma}}{[g(\tau)]^{1/n}} \right) + F(\tau) \right\}
\, ,
\end{align}
where various functions in $V_F$ are
\begin{align}
U(\tau,\sigma) = & 2\sigma - \gamma \, k(\tau) \, ,
\\
g(\tau) = & 1 + \frac{\epsilon}{\mu} \cosh \left( \frac{\tau}{2} \right) \, ,
\\
F(\tau) = & \epsilon^{4/3}\gamma \left[ K(\tau) \sinh \left( \frac{\tau}{2} \right)
\right]^2
\nonumber\\
& \times \left[ K(\tau)\cosh \left( \frac{\tau}{2} \right) - \frac{\epsilon/\mu}{
4\pi\epsilon^{4/3}\gamma g(\tau)} \right]^2 \, ,
\\\label{DefK}
K(\tau) = & \frac{[\sinh(2\tau)-2\tau]^{1/3}}{2^{1/3}\sinh\tau} \, .
\end{align}
One can check that (\ref{Vfts}) smoothly interpolates to the two-field potential derived
in Ref.~\cite{Baumann1} in the large $\tau$ limit $\epsilon^2\cosh\tau\approx
\epsilon^2e^\tau/2\approx r^3$, where $r$ is the usual radial coordinate of the singular
conifold\footnote{However, we have checked that once the volume modulus $\sigma$ is
stabilized in the adiabatic approximation we shall discuss next, there are deviations in
resultant single field potentials, due to different radial dependence of
$\sigma(\tau)$.}.

Having obtained the two-field $F$-term scalar potential $V_F(\tau,\sigma)$, the canonical
inflaton can be derived from the DBI action of a mobile D3 brane moving in the full
deformed conifold metric as the following integral expression;
\begin{equation}\label{DefCanInflaton}
\phi(\tau)=\sqrt{\frac{T_3}{6}}\epsilon^{2/3}\int_\tau \frac{d\tau '}{K(\tau')} \, .
\end{equation}
Here, we have used the explicit deformed conifold metric given in terms of radial and
angular coordinates (see, for example, Refs.~\cite{KS,Conenotes}), and one can see this
definition has the asymptotic limits
\begin{equation}\label{phiapprox}
\phi(\tau) \to \left\{
\begin{split}
& \sqrt{\frac{3}{2}T_3}r \, , & (\tau \gg 1)
\\
& \frac{\sqrt{T_3}}{2^{5/6}3^{1/6}}\epsilon^{2/3}\tau \, , & (\tau \ll 1)
\end{split}
\right.
\end{equation}
where we have used the definition $r^3=\epsilon^2\cosh\tau$ to rewrite the $\tau \gg
1$ limit. The expressions of the canonical inflaton in the large and small $\tau$
limits have been used in Refs.~\cite{Baumann1} and \cite{CGS1}, respectively.

As the deformed conifold throat is attached to a compact Calabi-Yau at some finite
ultraviolet radius $r_\mathrm{UV}$, it is important to stabilize the volume modulus
$\sigma$, which controls the overall size. Within the adiabatic approximation proposed in
Ref.~\cite{Baumann1}, such that $\sigma$ is stabilized at an instantaneous minimum as the
radial coordinate $\tau$ varies, this amounts to solving the equation
\begin{equation}\label{Defsigmastar}
\left. \frac{\partial(V_F+V_{\rm uplift})(\tau,\sigma)}{\partial\sigma}
\right|_{\sigma_\star[\phi(\tau)]} = 0 \, .
\end{equation}
Here, we have included the positive definite potential $V_{\rm
uplift}(\tau,\sigma)=(D_0+D_{\rm others})/[U(\tau,\sigma)]^2$, which is required to
uplift the total energy and to obtain a de Sitter phase. $V_{\rm uplift}(\tau,\sigma)$
can include the first term of $V_{D3\overline{D3}}(\phi)$ given by (\ref{VD3D3bar}) and
other supersymmetry breaking sources in the bulk as encoded in $D_{\rm
others}/[U(\tau,\sigma)]^2$, which can be generated by distant $\overline{\rm D3}$ or
wrapped D7 with supersymmetry breaking world volume fluxes~\cite{BKQ}\footnote{The
precise $U(\tau, \sigma)$ dependence in fact varies for different distant supersymmetry
breaking sources: for $\overline{\rm D3}$ the potential $\sim U(\tau,\sigma)^{-2}$ and
for D-term uplifting~\cite{BKQ} induced by D7 carrying supersymmetry breaking flux, it is
$\sim U(\tau,\sigma)^{-3}$. Here in the limit $U(\tau,\sigma)\gg 1$, we merely keep the
most dominant contribution.}. One can also parametrize the uplifting potential by
defining the uplifting ratio $s$ as
\begin{equation}
s = \frac{V_{\rm uplift}(0,\sigma_F)}{|V_F(0,\sigma_F)|} \, ,
\end{equation}
with $\sigma_F$ being given by $\partial_\sigma V_F(0,\sigma)|_{\sigma=\sigma_F}=0$.
The distant sources are essentially needed for a small positive cosmological
constant at the end of inflation after D3-$\overline{\rm D3}$ annihilation.
Combining this fact with the requirement that $s \lesssim 3$ during inflation for
avoiding runaway decompactification, one can deduce that $D_{\rm others}$ should
typically dominate over $D_{0}$. Alternatively, one can also argue that as the
distant sources are located in the unwarped region, it should naturally dominate
over the $\overline{D3}$ localized at the tip of the highly warped deformed
conifold~\cite{CGS1}.

Equation (\ref{Defsigmastar}) is transcendental and is usually solved numerically.
However to get a qualitative understanding, we can adopt a semi-analytic approach given
in Ref.~\cite{Baumann1}, where one sets the $\sigma$ dependence in $U(\tau,\sigma)$
equals to large fixed value $\sigma_0$ and treat (\ref{Defsigmastar}) as a quadratic
equation of the variable $\exp[-a\sigma_\star(\phi)]$. A double expansion in $1/\sigma_0$
and $\phi(\tau)$ around the tip region, such that $\phi(\tau)$ is approximated by the
$\tau \ll 1$ limit of (\ref{phiapprox}), then yields at leading order correction
\begin{equation}\label{sigmastar}
\sigma_\star(\phi) \approx \sigma_0 \left\{ 1 + \frac{1}{a\sigma_F} \left[
\frac{1}{3} + \frac{(2/3)^{2/3}\alpha}{8n(1+\alpha)\beta} \right]
\left(\frac{\phi}{\mpl}\right)^2 \right\} \, .
\end{equation}
In deriving the above expression we have used the approximation $a\sigma_0\approx
a\sigma_F+s/(a\sigma_F)$ given in Ref.~\cite{Baumann1}\footnote{Let us comment on
the difference between the expression for $\sigma_\star(\phi)$ in
Ref.~\cite{Baumann1}, which was given schematically by
$\sigma_\star(\phi)\approx\sigma_0[1+b_{3/2}(\phi/\mpl)^{3/2}]$, and our expression
(\ref{sigmastar}). In Ref.~\cite{Baumann1}, $V_F(\tau,\sigma)$ was calculated
exclusively for the large radius, singular conifold limit. The authors of
Ref.~\cite{Baumann1} then expanded in canonical inflaton $\phi\approx\sqrt{3T_3/2}r$
around the near tip region of deformed conifold to extract the radial dependence of
the stabilized volume. Here we improved upon such calculation, using
$V_F(\tau,\sigma)$ for the {\em entire} deformed conifold and expanding near the tip
of the deformed conifold using the small radius limit of the canonical inflaton
(\ref{phiapprox}) to obtain the expression (\ref{sigmastar}).}. Note that we have
introduced two important dimensionless parameters
\begin{align}
\label{alpha}
\alpha = & \frac{\epsilon}{\mu} \, ,
\\
\label{beta}
\beta = & \sqrt{\frac{T_3}{6}} \frac{\epsilon^{2/3}}{\mpl} \, .
\end{align}
Geometrically, $\alpha$ measures the depth which D7 branes extend into deformed
conifold, and $\beta$ is proportional to the warp factor $a_0$ at the tip. Of course
the analytic approximation for the stabilized volume $\sigma_\star(\phi)$ only gives
a qualitative understanding, and is expected to deviate from the actual behavior at
large radius. For full quantitative parameter scanning however, the numerical
solution to (\ref{Defsigmastar}) can also be readily implemented.

\begin{figure}[h]
 \begin{center}
 \includegraphics[width = 0.45 \textwidth]{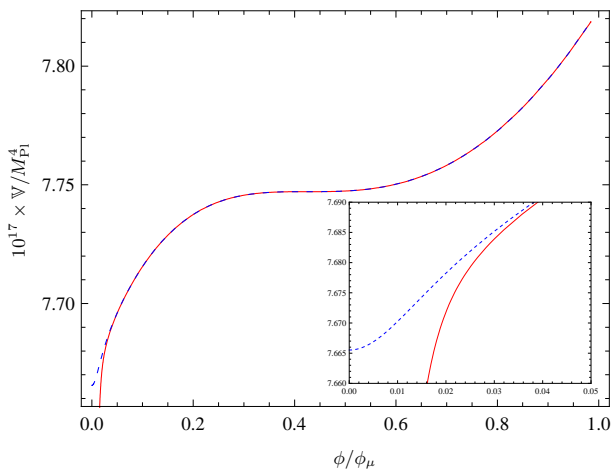}
 \end{center}
\caption{The plot of the potential (\ref{InflatonPotential}) as a function of
$\phi/\phi_\mu$, with $\phi_\mu^2 \equiv 3T_3/2 (2\mu^2)^{2/3}$. The parameters are set
the same as Case 1 of Table~\ref{table:local}. We show two extreme cases of $D(\phi)$:
either it is completely dominated by the Coulombic interaction (solid line) or by the
distant sources (dotted line). Note that the difference becomes only noticeable at the
region very close to the tip, as shown in the inset, which magnifies the potential in
this region. This implies inflation only ends when $\phi$ approaches close to the tip,
even if the potential is highly curved by the Coulombic term: in the case shown here,
$\phi_{\rm e} \approx 0.0105 \phi_\mu$, meanwhile the ``potential'' slow-roll parameter
with higher order corrections~\cite{Gong:2001he} gives $\phi_{\rm e} \approx 0.0700
\phi_\mu$. Note that $|\eta| \equiv |\mpl^2 \mathbb{V}''/\mathbb{V}| = 1$ well before
this point, at $\phi \approx 0.324 \phi_\mu$.}
 \label{figure:potential}
\end{figure}

Combining our expression for the stabilized volume $\sigma_\star(\phi)$ given by
(\ref{sigmastar}), the potential $V_{D3\overline{D3}}(\phi)+V_{\rm stab.}(\phi)$ for
the D7 embedding (\ref{Kuperembedding}) in the entire deformed conifold is finally
given by
\begin{widetext}
\begin{align}
\label{InflatonPotential}
V_{D3\overline{D3}}(\phi) + V_{\rm stab.}(\phi) = &
\frac{2a^2\kappa^2|A_0|^2e^{-2a\sigma_\star(\tau)}}{\{U[\tau,\sigma_\star(\tau)]\}^2}
|g(\tau)|^{2/n} \left\{\frac{U[\tau,\sigma_\star(\tau)]}{6} + \frac{1}{a} \left( 1 -
\frac{|W_0|}{|A_0|} \frac{e^{a\sigma_\star(\tau)}}{[g(\tau)]^{1/n}} \right) + F(\tau)
\right\} + \frac{D(\phi)}{\{U[\tau,\sigma_\star(\tau)]\}^2} \, ,
\\
\label{DefDphi}
D(\phi) = & D_0 \left( 1 - \frac{27D_0}{64\pi^2\phi^4} \right) + D_{\rm others} \, .
\end{align}
\end{widetext}
Here, we should regard the radial coordinate $\tau$ to be an implicit function of the
canonical inflaton $\phi$ given by (\ref{DefCanInflaton}). In addition, as shown in
Ref.~\cite{CGS1} the residual angular isometry directions becomes degenerate along the
trajectory (\ref{StabTraj}). Therefore D3-$\overline{\rm D3}$ separation ${|\bf
y-\bar{y}|}$ is purely radial and proportional to the canonical inflaton $\phi(\tau)$ for
the entire deformed conifold. In Fig.~\ref{figure:potential}, we show the plot of the
potential (\ref{InflatonPotential}) with the parameters given by Case 1 of
Table~\ref{table:local}.

Let us conclude this section by revisiting the $\eta$ problem discussed in
Ref.~\cite{Baumann1}, now with the potential (\ref{InflatonPotential}) valid in the
region near the tip with the small field canonical inflaton given in
(\ref{phiapprox}) and the stabilized volume (\ref{sigmastar}). After some
expansions, we can obtain
\begin{equation}\label{DefVeta}
\frac{V_{D3\overline{D3}}(\phi)+V_{\rm stab.}(\phi)}{V_{D3\overline{D3}}(0)+V_{\rm
stab.}(0)} \approx 1 + \frac{\phi^2}{3\mpl^2} \left[ 1 + \mathcal{O}\left(
\frac{1}{\sigma_0} \right) \right] + \mathcal{O}(\phi^4) \, .
\end{equation}
Notice that the dependence of the gaugino condensate on the mobile D3 brane position {\it
does} give corrections to the inflaton mass in the near tip region. However, such
corrections are suppressed by the large stabilized volume $\sigma_0$, and are
insufficient to give small inflaton mass. Thus, $\eta$ remains of order
one\footnote{Notice that on the other hand $|\varepsilon|\ll 1$, as we do not have
trans-Planckian field displacement $\Delta\phi/\mpl \ll 1$.}. This is in fact consistent
with the analysis in Ref.~\cite{Baumann1} using the singular conifold approximation, that
the inflection point $\eta = 0$ only appears at some intermediate radius\footnote{Notice
that the analysis here is accurate for the near tip region. The required cancelation term
$\propto\phi^{3/2}$ for obtaining inflection point, only appears when the large radius
canonical inflaton (\ref{phiapprox}) and the associated stabilized volume expression are
substituted in the derivation.}.

\subsection{Parametrization of the bulk effects}

To account for the ultraviolet physics arising from attaching the warped throat to a
compact Calabi-Yau, a useful parametrization of the leading corrections was given in
Ref.~\cite{Baumann2}. The authors employed gauge/string correspondence for the warped
deformed conifold (see, for example, Refs.~\cite{KS,Conenotes}), where the position of
the mobile D3 is identified with the Coulomb branch vacuum expectation value of the dual
field theory. In such a holographic formulation, the symmetry breaking bulk effects can
be encoded by coupling a field theory operator ${\mathcal{O}}_{\Delta}$ of scaling
dimension $\Delta$ to its dual bulk mode and a perturbation to inflaton $\phi$ potential
is generated as
\begin{equation}\label{bulkDeltaV}
\Delta V = -c_{\Delta} a_0^4 T_3\left(\frac{\phi}{\phi_\mathrm{UV}}\right)^{\Delta}
\, .
\end{equation}
Here $\phi_\mathrm{UV}=\sqrt{3T_3/2}r_{\rm UV}$ and $r_{\rm UV}$ is the radius at which
the deformed conifold throat joins the compact Calabi-Yau. The positive constant
$c_{\Delta}$ depends on the specific distant fluxes or brane configurations\footnote{To
be specific, $c_{\Delta}$ used here only incorporate strictly bulk effects. This is in
contrast with Ref.~\cite{Baumann2}, where the $c_{\Delta}$ coefficients there can receive
{\em both} local and bulk contributions.}. Varying its value allows us to parametrize our
ignorance about this information. The normalization of $a_0^4 T_3$ in (\ref{bulkDeltaV})
comes from the estimated energy required to move the mobile D3 from its supersymmetric
minimum to the four cycle moduli stabilizing D7 wraps. This is proportional to the height
of the anti de Sitter potential barrier which in our more detailed setup should be
identified explicitly with $|V_{F}(0,\sigma_F)|$.

There can in fact be whole series of perturbations of the form given in
(\ref{bulkDeltaV}). However the two leading contributions come from the lowest
chiral multiplet of dimension $3/2$, $\mathcal{O}_{3/2}$, and the lowest non-chiral
multiplet of dimension $2$, $\mathcal{O}_2$. For our case, the bulk potential as
denoted in (\ref{TotalV}) is then given by
\begin{equation}\label{Vbulk}
V_{\rm bulk}(\phi) = -|V_{F}(0,\sigma_F)| \left[ c_{3/2} \left(
\frac{\phi}{\phi_{\rm UV}} \right)^{3/2} + c_2 \left( \frac{\phi}{\phi_{\rm UV}}
\right)^{2} \right] \, .
\end{equation}
One can of course include other higher dimensional operators in $V_{\rm bulk}(\phi)$. The
terms above are merely to illustrate the importance of bulk physics in our later sample
parameter scannings\footnote{In Ref.~\cite{Panda}, an earlier attempt to perform
parameter scanning using (\ref{Vbulk}) is given.}. However, one should note that when
more than one $(\phi/\phi_{\rm UV})^{\Delta}$ is turned on in (\ref{Vbulk}), there are
generally additional angular perturbations. This comes from the fact that individual
coefficient $c_{\Delta}$ is obtained from integrating out complicated angular
dependences. When more than one $c_{\Delta}$ are involved, it is generally not possible
to perform such integrating out\footnote{We are grateful to Daniel Baumann for pointing
this out to us.}.

\section{Constraining parameter space: microscopic and observational}

\subsection{Microscopic constraints}

Let us first list out the explicit parameters specifying the total single field inflaton
potential ${\mathbb{V}}(\phi)=V_{D3\overline{D3}}[\phi,\sigma_\star(\phi)]+V_{\rm
stab.}[\phi,\sigma_\star(\phi)]+V_{\rm bulk}[\phi,\sigma_\star(\phi)]$; they are
\begin{equation}\label{localindepparameters}
\left\{n\,, |A_0|\,, |W_0|\,, s\,, \epsilon\,, \mu\,, c_{3/2}\,, c_2 \right\} \, .
\end{equation}
Here, we have used the $F$-term flatness condition $D_{\sigma}W|_{\sigma_F}=0$,
\begin{equation}\label{defsigmaF}
e^{a\sigma_F} = \frac{|A_0|}{|W_0|} \left( 1 + \frac{2}{3}a\sigma_F \right) \left( 1
+ \alpha \right)^{1/n} \, ,
\end{equation}
to exchange $|W_0|$ for $\sigma_F$. From the perspective of K\"ahler moduli
stabilization, $\sigma_F$ should be regarded as a derived parameter, which is obtained as
soon as the hierarchy between $|A_0|$ and $|W_0|$ is specified\footnote{Furthermore, in
this paper, we shall consider the configuration where moduli stabilizing D7 brane is
sufficiently far away from the tip of the deformed conifold. Therefore the the term $1 +
\alpha$ with $\alpha = \epsilon/\mu \ll 1$ in (\ref{defsigmaF}) only gives insignificant
shift in $\sigma_F$.}. Before comparing with the observational data, there are additional
microscopic requirements that need to be satisfied a priori. Here, we list them below.

\begin{itemize}

\item
The string coupling $g_s$ should be small, i.e. $g_s\ll 1$ to ignore the string loop
corrections to the supergravity action. The physical radius of the three sphere at the
tip of deformed conifold is $g_s M\alpha'$; thus, we also need $g_s M\gg 1$~\cite{KS}.

\item
The ultraviolet cutoff $r_\mathrm{UV}$ should be large such that $r_\mathrm{UV}/l_s
\gg 1$ for valid supergravity solution. This sets the upper bound on the
displacement for $\phi$, hence the total number of $e$-folds. Moreover, the unit of
five form flux $N=KM$ controlling the size of conifold needs to be large for the
supergravity approximation to be valid. These geometric requirements combine to give
a strong bound on the tensor-to-scalar ratio $r$~\cite{tensor-to-scalar},
\begin{equation}\label{DefMp}
\frac{4}{N} \gtrsim \left( \frac{\phi_\mathrm{UV}}{\mpl} \right)^2 \gtrsim 100
\times r \, ,
\end{equation}
where the $\gtrsim$ sign is to indicate that bulk volume can also give significant
contribution to $V_6^{w}$. This can be obtained from the relation between the four
dimensional reduced Planck mass $\mpl$ and the warped volume $V_6^{w}$ of the compact six
manifold. Given $N\gg 1$, the inequality (\ref{DefMp}) implies that warped brane
inflation yields negligible tensor-to-scalar ratio.

\item
The stabilized volume modulus $\sigma_F$ should also be at large values for the
$\alpha'$ corrections to be suppressed. This can be ensured by tuning the bulk flux
to generate a large hierarchy between $|A_0|$ and $|W_0|$, i.e. $|A_0|/|W_0|\gg 1$,
since $a=2\pi/n$ is typically smaller than 1 so that large $\sigma_F$ can be readily
produced.
To avoid the backreaction of D7 branes on the deformed conifold, however, $n$
should also be such that $n/M\ll 1$. This ensures that the resultant geometry is
smooth at the end of duality cascade, rather than cascading into singular conifold
throat.

\item
Finally, the uplifting ratio $s$ is bounded within the range $1 \le s \le
\mathcal{O}(3)$ to ensure a small positive cosmological constant at the end of
inflation. The upper bound here arises from preventing runaway decompactification.
Such requirement effectively couples the scale of $|V_F|$ and the scale of the
uplifting term(s) $V_{\rm uplift}(\phi)$.

\end{itemize}

\subsection{Comparison with observations}

In this section we shall first consider some generic features of the inflaton potential
(\ref{TotalV}) with $c_{3/2}=c_2=0$, i.e. involving only
$V_{D3\overline{D3}}(\phi)+V_{\rm stab.}(\phi)$ given by (\ref{InflatonPotential}). In
particular, we discuss which parameters listed in (\ref{localindepparameters}) have more
impact on the overall scale or the detailed shape of the inflaton potential, as this is
useful for an efficient full parameter space scanning. Next, we shall present some sample
parameter sets to demonstrate that $V_{D3\overline{D3}}(\phi)+V_{\rm stab.}(\phi)$ can
indeed yield observationally consistent results. Such scanning for our complete potential
is in line with the existing literature~\cite{branescan,ClineScan}. Finally, we shall
scan the perturbations due to $V_{\rm bulk}(\phi)$ on these observationally consistent
local potential $V_{D3\overline{D3}}(\phi)+V_{\rm stab.}(\phi)$, and demonstrate that
bulk contributions generically need to be highly fine-tuned to preserve such results.

Let us first consider the amplitude of the power spectrum of the curvature
perturbation $\mathcal{P}_\mathcal{R}$ and the corresponding spectral index
$n_\mathcal{R}$, which are tightly constrained by recent cosmological
observations~\cite{Tegmark:2006az,Komatsu:2008hk}. On the largest observable scales
the slow-roll approximation holds at a good enough accuracy (see later discussion),
we can express them as
\begin{align}
\label{obsP}
\mathcal{P}_\mathcal{R} = & \frac{\mathbb{V}}{24\pi^2\varepsilon\mpl^4} = (2.41 \pm
0.22) \times 10^{-9} \, ,
\\
\label{obsn}
n_\mathcal{R} = & 1 - 6\varepsilon + 2\eta = 0.963 \pm 0.028 \, ,
\end{align}
at 95\% confidence level. Here, (\ref{obsP}) and (\ref{obsn}) are evaluated at $\phi_{\rm
CMB}$, the value of the canonical inflaton at the CMB scale, and should be determined by
integrating backwards $60$ $e$-folds\footnote{There exists some level of uncertainty on
exactly when the perturbation on the largest observable scales is generated. Depending on
the detail of the model, the corresponding $e$-fold is supposed to lie between 50 and
60~\cite{efold_estimate}. But provided that the curvature power spectrum is nearly scale
invariant it does not cause too significant differences. Thus in the remaining text we
evaluate $\mathcal{P}_\mathcal{R}$ and $n_\mathcal{R}$ at 60 $e$-folds before the end of
inflation.} from the end of inflation. The inflationary scale is expected to be
approximately constant around the CMB scale, and, in particular, for our model, it is
expected to occur near the ``inflection point'' where the majority of $e$-folds is
generated. Explicitly, the combination
\begin{equation}\label{VMp4}
(s-1)|V_F(0,\sigma_F)|
\approx(s-1)\frac{a^2|A_0|^2e^{-2a\sigma_F}}{3M_p^2(2\sigma_F)}\approx
{\mathbb{V}}(\phi_{\rm CMB})
\end{equation}
largely sets the overall scale of inflation in our model. The deviation from (\ref{VMp4})
due to the motion of mobile D3 is essentially a small fluctuation around it. If the
energy associated with the inflaton is too large, this would in fact lead to runaway
decompactification~\cite{KLbound}. The slow-roll parameter $\varepsilon$ is also small
around the CMB scale, but it varies more rapidly than $\mathbb{V}(\phi)$. We therefore
conclude to obtain an observationally consistent value of (\ref{obsP}), it is easier to
fix the combination (\ref{VMp4}) which sets the overall scale, then vary other parameters
such as $\epsilon$ and $\mu$, which affect the shape of $\mathbb{V}(\phi)$ around
$\phi_{\rm CMB}$.

It is also worth noting that while the uplifting ratio $s$ or $V_{\rm uplift}(\phi)$
is fixed, one can still vary the ratio between the distant uplifting $(\propto
D_{\rm others})$ and contribution from $\overline{\rm D3}$ at the tip of $(\propto
D_0)$. This also varies the D3-$\overline{\rm D3}$ Coulombic attraction in
(\ref{VD3D3bar}). However, as such highly warped attraction only becomes significant
near the tip region, it is important to use the full scalar potential
(\ref{InflatonPotential}) to study any change in the trajectory. Furthermore, at the
relatively large distance where the CMB scale lies, the Coulombic attraction is
effectively absent. The variation of $D_0/D_{\rm others}$ therefore should not
affect significantly the observational predictions\footnote{It is in principle
possible to finely tune the CMB scale to small radius~\cite{ClineScan}, but there
one should again use the full potential valid for that region
(\ref{InflatonPotential}) to study the effects of varying $D_0/D_{\rm others}$ on
the trajectory.}. This is indeed the case as illustrated in Fig.~\ref{figure:potential}
and Table~\ref{table:local}.

Now we would like to present some sample parameter scannings for
$V_{D3\overline{D3}}(\phi)+V_{\rm stab.}(\phi)$. The strategy is that we shall
further systematically fix the parameters $n, |A_0|, |W_0|$ and $s$ to some
appropriate fiducial values by hand. This allows us to roughly fix the overall scale
of the inflaton potential following (\ref{VMp4}). We then generate a range of
observationally consistent parameter sets by scanning in $\epsilon$-$\mu$ or
equivalently $\alpha$-$\beta$ plane.

Let us briefly describe how the fiducial values for these other parameters are chosen.
The number of probe D7s $n$ can first be fixed to be sufficiently small. This is because
$n$ appears mostly with $\sigma_\star$ or in $[g(\tau)]^{1/n}$. With $\alpha
\cosh(\tau/2)< 1$ and $\sigma_F\gg 1$, the dependence of the inflaton potential on $n$ is
insignificant comparing with other parameters. To fix the values of $|A_0|, |W_0|$ and
$s$, as mentioned earlier that their relative sizes are fixed by compactification
constraints (\ref{defsigmaF}), we need to set the ratio $|A_0|/|W_0|$ large to ensure the
volume modulus is fixed at large value $\sigma_F$. For the actual value of $|A_0|$, we
note that as $|A_0|$ is related to the dynamical scale $\Lambda$ at which gaugino
condensation takes place~\cite{ClineScan}, therefore it is necessary to have
$|A_0|^{1/3}\sim \Lambda\le \mpl$. To fix the uplifting ratio $s$, the resultant
cosmological constant should be small and positive at the end of inflation, but not
necessarily at our current value as, for example, there can be further dynamical
processes, e.g. topological changes after inflation, which can change its value. The
specific numerical values for $\{n\,, |A_0|\,, |W_0|\,, s\}$ used in our scanning are
given in Table ~\ref{table:local}.

With the full inflaton potential given by (\ref{InflatonPotential}) and
(\ref{Vbulk}), we can {\em exactly} solve the system and subsequently identify where
inflation ends, i.e. $\varepsilon = 1$. This is most easily done by solving, instead
of the Friedmann equation, the Hamilton-Jacobi equation
\begin{equation}\label{HJeq}
2\mpl^4[H'(\phi)]^2 - 3\mpl^2[H(\phi)]^2 + \mathbb{V}(\phi) = 0 \, .
\end{equation}
We can thus calculate the {\em exact} number of $e$-folds $\mathcal{N}_e$ given by
\begin{equation}
\mathcal{N}_e(\phi) = \mpl^{-1} \int_{\phi_\mathrm{e}}^\phi
\frac{d\phi}{\sqrt{2\varepsilon}} \, ,
\end{equation}
with $\varepsilon$ defined as (\ref{SRepsilon}) and $\phi_\mathrm{e}$ given by solving
(\ref{HJeq}), and subsequently identify $\phi_\mathrm{CMB}$ where
$\mathcal{N}_e(\phi_\mathrm{CMB}) = 60$. Note that for $\phi_{\rm e}$, we explicitly
consider two limiting cases, where the uplifting is exclusively by the distant sources or
by the warped $\overline{\rm D3}$. As mentioned earlier and checked in our scannings that
$\varepsilon\ll 1$ until the tip of deformed conifold for distant uplifting, thus
$\phi_{\rm e}=0$ in this case. Whereas for warped $\overline{\rm D3}$, $\phi_{\rm e}$ can
also be determined at a small radius by solving (\ref{SRepsilon}). Essentially we expect
that the end point $\phi_{\rm e}$ will vary continuously as we dial between the two limit
cases. Furthermore, at any viable $\phi_\mathrm{CMB}$ the potential is very flat so that
$|\eta| \ll 1$, we therefore make use of the simplified slow-roll formulae (\ref{obsP})
and (\ref{obsn}) to estimate $\mathcal{P}_\mathcal{R}$ and $n_\mathcal{R}$ respectively,
instead of solving the perturbation equations mode by mode.

\begin{figure*}[t]
 \begin{center}
 \includegraphics{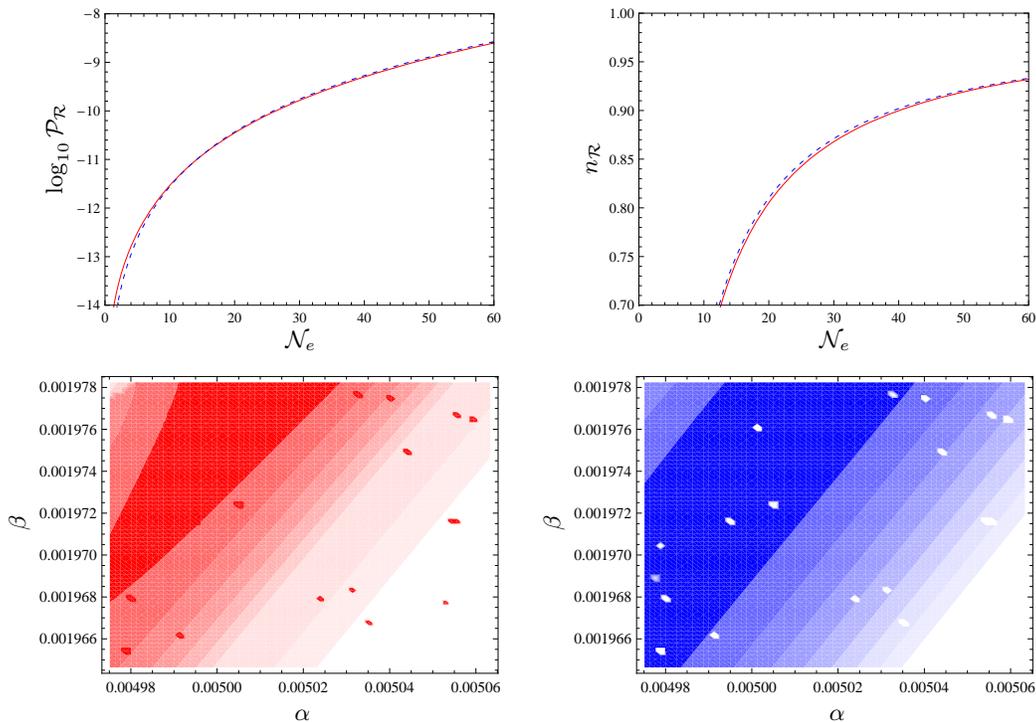}
 \end{center}
\caption{(Upper panels) the plots of (left panel) $\log_{10}\mathcal{P}_\mathcal{R}$ and
(right panel) $n_\mathcal{R}$ as functions of $\mathcal{N}_e$, and (lower panels) the
contour plots of (left panel) $\mathcal{P}_\mathcal{R}$ and (right panel) $n_\mathcal{R}$
in the $\alpha$-$\beta$ plane for Case 1 given in Table~\ref{table:local}. In the upper
panels we show the two extreme cases where $D(\phi)$ given by (\ref{DefDphi}) is
completely dominated by either the Coulombic interaction (solid line) or the distant
sources (dotted line). Meanwhile, in the lower panels we only present the case with the
distant sources completely dominating. In the contour plot of $\mathcal{P}_\mathcal{R}$,
the contours denote 2.5, 2.6, 2.7, 2.8, 2.9, 3.0, 3.5, $4.0 \times 10^{-9}$ from the
innermost line. Likewise we have set 0.9325, 0.9300, 0.9275, 0.9250, 0.9225, 0.9200,
0.9175, 0.9150 for the contour plot of $n_\mathcal{R}$. The dots in the contour plots are
numerical glitches. We have obtained qualitatively the same contour plots when the
Coulombic interaction is dominating instead, with the deep colored region a bit enlarged
($\mathcal{P}_\mathcal{R}$) and shrunk ($n_\mathcal{R}$).}
 \label{figure:plots}
\end{figure*}

In Table~\ref{table:local}, we present three sets of $\alpha$ and $\beta$, which give
similar predictions on $\mathcal{P}_\mathcal{R}$ and $n_\mathcal{R}$ for
$V_{D3\overline{D3}}(\phi)+V_\mathrm{stab.}(\phi)$. The values of $W_0$ and $A_0$ are the
same in both Cases 1 and 2. The scanned results suggest that locally there exists a
region of degeneracies in $\alpha$-$\beta$ plane with the other parameters fixed, as
explicitly demonstrated in the lower panels of Fig.~~\ref{figure:plots}. Furthermore, if
the other parameters are allowed to vary, we can produce similar prediction in an even
wider range of parameter sets. A sample parameter set with different $|A_0|$ and $|W_0|$
is presented as Case 3 in Table~\ref{table:local}. Note that from
Fig.~\ref{figure:plots}, a fractional change of $\mathcal{O}(1)$\% in either $\alpha$ or
$\beta$ can easily move the values of $\mathcal{P}_\mathcal{R}$ and $n_\mathcal{R}$ to
observationally inconsistent regimes. Here, we have scanned only the vicinity of a given
$\{\alpha\,,\beta\}$, and it is not entirely clear (although suggestive) such
$\mathcal{O}(1) \%$ tuning in $\alpha$-$\beta$ plane holds for a wider range. It would be
interesting to return to this issue in a more complete scanning in the future.

\begin{table}
\begin{tabular}{c|cc|cc|cc}
  \hline\hline
  & $|W_0|$ & $|A_0|$ & $\alpha$ & $\beta$ &
  $\mathcal{P}_\mathcal{R} \times 10^9$ & $n_\mathcal{R}$
  \\
  \hline
  Case 1 & $2.92485\times10^{-6}$ & 0.0085 & 1/200 & 1/508 & 2.66644 & 0.933109
  \\
  &  & & & & 2.49420 & 0.932009
  \\
  \hline
  Case 2 & $2.92485\times10^{-6}$ & 0.0085 & 1/100 & 1/320 & 2.59208 & 0.934267
  \\
  & & & & & 2.42615 & 0.933175
  \\
  \hline
  Case 3 & $3.3 \times 10^{-6}$ & 0.066 & 1/100 & 1/350 & 2.36186 & 0.934743
  \\
  & & & & & 2.19847 & 0.933838
  \\
  \hline\hline
\end{tabular}
\caption{Three sets of parameters that give the viable values of
$\mathcal{P}_\mathcal{R}$ and $n_\mathcal{R}$. We have fixed $n = 8$ and $s = 1.07535$
for all the cases. The values of the tensor-to-scalar ratio $r = 16\varepsilon$ and the
non-linear parameter $f_\mathrm{NL} = \mathcal{O}(\varepsilon,\eta)$ are unobservably
small and hence we do not present them here. The first line of each case corresponds to
the complete domination of the distant sources, while the second to that of the Coulombic
interaction. Note that as shown in the first two cases, with a given set of $n$, $s$,
$|W_0|$ and $|A_0|$, a different combination of $\alpha$ and $\beta$ yields similar
values of $\mathcal{P}_\mathcal{R}$ and $n_\mathcal{R}$. Also, in the last case with
another set of $|W_0|$, $|A_0|$, $\alpha$ and $\beta$ we can find observationally
consistent values of $\mathcal{P}_\mathcal{R}$ and $n_\mathcal{R}$.}
 \label{table:local}
\end{table}

\subsection{Bulk effects scanning}

Having presented a range of the observationally consistent parameter sets for
$V_{D3\overline{D3}}(\phi)+V_{\rm stab.}(\phi)$, we shall now consider the perturbations
on them, due to the unknown bulk physics parametrized by $V_{\rm bulk}(\phi)$
(\ref{Vbulk}). In particular, for the local inflection point based inflationary
trajectories, we shall perform a sample scanning in the $c_{3/2}$-$c_2$ plane to
demonstrate that they typically need to be of order $10^{-8}$-$10^{-9}$ to preserve
consistent observational predictions.

For numerical purpose, we slightly recast (\ref{Vbulk}) as
\begin{align}\label{Vbulk2}
V_\mathrm{bulk}(\phi) = & -|V_{F}(0,\sigma_F)|
\nonumber\\
& \times \left\{ c_{3/2}' \alpha \left[ \int_\tau \frac{d\tau'}{K(\tau')}
\right]^{3/2} + c_2' \alpha^{4/3} \left[ \int_\tau \frac{d\tau'}{K(\tau')} \right]^2
\right\} \, .
\end{align}
In the above we have used
\begin{equation}
\frac{\phi}{\phi_\mathrm{UV}} = \frac{\phi_\mu/\phi_\mathrm{UV}}{2^{1/3}\cdot3}
\alpha^{2/3} \int_\tau \frac{d\tau'}{K(\tau')} \, ,
\end{equation}
with $\phi_\mu/\phi_\mathrm{UV} \lesssim 1$ being a number (This is denoted as
$Q_\mu^{-1}$ in Refs.~\cite{Baumann1,CGS1}) and various order one numerical factors are
absorbed into a newly defined constant $c_\Delta'$. Hereafter, we shall drop this prime
notation. In general, as $V_{D3\overline{D3}}(\phi)$ yields a delicate inflection point
based inflation, we expect the value of $c_\Delta$ needs to be finely tuned. In
Table~\ref{table:bulk} we show a summary of the effects of $V_\mathrm{bulk}$.

\begin{table*}
\begin{tabular}{c|c|c|cccc|cccc}
  \hline\hline
  \multicolumn{3}{c|}{} & \multicolumn{4}{c|}{$c_{3/2}$} & \multicolumn{4}{c}{$c_2$}
  \\
  \cline{4-11}
  \multicolumn{3}{c|}{} & $10^{-9}$ & $10^{-8}$ & $10^{-7}$ & $10^{-6}$ &
                          $10^{-9}$ & $10^{-8}$ & $10^{-7}$ & $10^{-6}$
  \\
  \hline
  Case 1 & Distant & $\mathcal{P}_\mathcal{R} \times 10^9$
  & 2.71386 & 3.17635 & 13.7483 & 26670.1
  & 2.74701 & 3.58118 & 37.8218 & 0.0559217$^*$
  \\
  & sources & $n_\mathcal{R}$
  & 0.932540 & 0.927506 & 0.883792 & 0.657621
  & 0.932149 & 0.923750 & 0.856258 & 0.552413
  \\
  \cline{2-11}
  & Coulomb & $\mathcal{P}_\mathcal{R} \times 10^9$
  & 2.53682 & 2.95080 & 11.9644 & 11348.2
  & 2.56657 & 3.31098 & 31.2301 & 0.0138777$^*$
  \\
  & interaction & $n_\mathcal{R}$
  & 0.931448 & 0.926480 & 0.883233 & 0.657668
  & 0.931062 & 0.922766 & 0.855908 & 0.552458
  \\
  \hline
  Case 2 & Distant & $\mathcal{P}_\mathcal{R} \times 10^9$
  & 2.63754 & 3.08041 & 13.0733 & 22761.8
  & 2.66847 & 3.45614 & 34.6474 & 0.0399836$^*$
  \\
  & sources & $n_\mathcal{R}$
  & 0.933704 & 0.928724 & 0.885438 & 0.661093
  & 0.933327 & 0.925100 & 0.858794 & 0.559159
  \\
  \cline{2-11}
  & Coulomb & $\mathcal{P}_\mathcal{R} \times 10^9$
  & 2.46750 & 2.86903 & 11.5801 & 10664.9
  & 2.49564 & 3.20800 & 29.4348 & 0.01153290$^*$
  \\
  & interaction & $n_\mathcal{R}$
  & 0.932613 & 0.927646 & 0.884425 & 0.659278
  & 0.932238 & 0.924026 & 0.857777 & 0.556756
  \\
  \hline
  Case 3 & Distant & $\mathcal{P}_\mathcal{R} \times 10^9$
  & 2.40944 & 2.87842 & 14.8411 & 70552.3
  & 2.44157 & 3.27915 & 43.8420 & 0.186355$^*$
  \\
  & sources & $n_\mathcal{R}$
  & 0.934097 & 0.928393 & 0.879688 & 0.636830
  & 0.933668 & 0.924284 & 0.850560 & 0.528232
  \\
  \cline{2-11}
  & Coulomb & $\mathcal{P}_\mathcal{R} \times 10^9$
  & 2.24107 & 2.65907 & 12.7450 & 27115.1
  & 2.26982 & 3.01367 & 35.5661 & 0.0400323$^*$
  \\
  & interaction & $n_\mathcal{R}$
  & 0.933199 & 0.927564 & 0.879325 & 0.636940
  & 0.932776 & 0.923498 & 0.850383 & 0.528320
  \\
  \hline\hline
\end{tabular}
\caption{The effects of the bulk terms for each case of Table~\ref{table:local}. For
definiteness, we have turned on either $c_{3/2}$ or $c_2$, not both of them at the same
time. This was also needed to ensure that we can avoid additional angular perturbations
mentioned earlier in the main text. Also note that the values of
$\mathcal{P}_\mathcal{R}$ when $c_2 = 10^{-6}$, denoted by a superscript $*$ in the last
column, are bare ones and the factor of $10^9$ is {\em not} multiplied.}
 \label{table:bulk}
\end{table*}

Specifically, from Table~\ref{table:bulk}, we can see that a very slight disturbance of
the bulk effects of magnitude $10^{-9}$-$10^{-8}$ for both $c_{3/2}$ and $c_2$ can push
the otherwise viable predictions into the regions beyond 2-$\sigma$ errors. As
$V_\mathrm{bulk}$ is negative definite, it pushes down the inflaton potential further so
that the previously flat region becomes flatter or is even changed into a local minimum.
Naturally the amplitude of $\mathcal{P}_\mathcal{R}$ increases, while $n_\mathcal{R}$
deviates further from 1 as the value of the coefficients $c_{3/2}$ and $c_2$ get larger.
These tendencies are clearly shown in Table~\ref{table:bulk}. Occasionally $V_{\rm bulk}$
can improve the relevant predictions to be closer to the current observations. For
example, in Case 3, the bulk terms move the value of $\mathcal{P}_\mathcal{R}$ to the
central value of the observationally allowed region and leave $n_\mathcal{R}$ more or
less the same with small $c_{3/2}$ and $c_2$. One may thus hope that by adding $V_{\rm
bulk}(\phi)$ to an unviable $V_{D3\overline{D3}}(\phi)+V_{\rm stab.}(\phi)$,
observationally consistent results can be obtained. However, we expect in general
$c_{3/2}$ or $c_{2}$ need to be of order $10^{-8}$-$10^{-9}$  to achieve such objective.

\section{Discussions}

In this paper, we have discussed in detail the inflaton potential governing the motion of
a mobile D3 in the entire warped deformed conifold. In particular, we have included both
the effects of moduli stabilization and other bulk physics. We then have performed some
sample scannings to demonstrate that without the bulk perturbations, there can be
significant degeneracies in the conifold deformation parameter $\epsilon$ and the D7
embedding parameter $\mu$ for producing observationally consistent predictions. However,
as the bulk perturbations are included, we have explicitly shown that their magnitudes
need to be $10^{-8}$-$10^{-9}$ to preserve the observationally consistent parameter
sets\footnote{An obviously interesting question would be whether the smallness of bulk
perturbation coefficients $c_{\Delta}$ really constitutes a significant fine-tuning, or
they are just tied to the choice of having inflection point inflation in the throat. To
answer this question fully, we believe it requires better than our current understanding
of UV physics and beyond the scope of investigations here.}. The results presented here
provide the beginning systematic steps towards a complete brane scanning in the warped
throat, and, in particular, highlight the importance of the bulk effects.

It would be very interesting to follow the steps outlined here and perform a full
scanning over the parameters listed in (\ref{localindepparameters}). This clearly
requires intensive computational undertakings. However given the rich parameter space and
the degeneracies we have shown in the sample scannings, barring the observation of the
primordial gravitational waves, it is likely that there remain significant regions in the
parameter space for the warped brane inflation to match the future data. Moreover, a
variant of the inflation model presented here is recently proposed in Ref.~\cite{CHS}. In
such a construction the gravitino mass $m_{3/2}$ can be made smaller than the Hubble
scale $H$, hence circumventing the phenomenological bound given in Ref.~\cite{KLbound}.
It would clearly be interesting to generalize the analysis here and scan the parameter
space for such variant, and search for an explicit example of a parameter set that gives
TeV scale gravitino mass and observationally consistent cosmological predictions.

\subsection*{Acknowledgement}

We thank Gary Shiu for collaboration and discussions at the early stage of this project.
We are also grateful to Ana Ach\'ucarro, Daniel Baumann, James Cline, Shamit Kachru,
Gonzalo Palma, Fernando Quevedo, Koenraad Schalm and Bret Underwood for comments and
suggestions. HYC appreciates the hospitality of Stanford Institute for Theoretical
Physics, where part of this work was conducted. The work of HYC is supported in part by
NSF CAREER Award No. PHY-0348093, DOE grant DE-FG-02-95ER40896, a Research Innovation
Award and a Cottrell Scholar Award from Research Corporation, and a Vilas Associate Award
from the University of Wisconsin. JG is partly supported by the Korea Research Foundation
Grant KRF-2007-357-C00014 funded by the Korean Government at the early stage of this
work, and is currently supported in part by a VIDI and a VICI Innovative Research
Incentive Grant from the Netherlands Organisation for Scientific Research (NWO).

\end{document}